\documentclass[sigconf]{acmart}

\usepackage{booktabs} 
\usepackage{enumitem}
\usepackage{subcaption}
\usepackage[utf8]{inputenc}

\setcopyright{rightsretained}

\acmDOI{}

\acmISBN{}

\acmConference[arXiv]{arXiv}{2017}{https://arxiv.org/} 
\acmYear{2017}
\copyrightyear{2017}

\acmPrice{15.00}

\begin{document}
\title{Probabilistic Perspectives on Collecting Human Uncertainty\\ in Predictive Data Mining}
\titlenote{Full dataset and evaluation routines available at https://jasbergk.wixsite.com/research}

\author{Kevin Jasberg}
\affiliation{%
  \institution{Web Science Group\\Heinrich-Heine-University Duesseldorf}
  \city{Duesseldorf} 
  \state{Germany} 
  \postcode{45225}
}
\email{kevin.jasberg@uni-duesseldorf.de}

\author{Sergej Sizov}
\affiliation{%
  \institution{Web Science Group\\Heinrich-Heine-University Duesseldorf}
  \city{Duesseldorf} 
  \state{Germany} 
  \postcode{45225}
}
\email{sizov@hhu.de}

\renewcommand{\shortauthors}{}

\pagenumbering{arabic}
\pagestyle{plain}
\thispagestyle{plain}
\begin{abstract}
In many areas of data mining, data is collected from humans beings. In this contribution, we ask the question of how people actually respond to ordinal scales. The main problem observed is that users tend to be volatile in their choices, i.e. complex cognitions do not always lead to the same decisions, but to distributions of possible decision outputs. This human uncertainty may sometimes have quite an impact on common data mining approaches and thus, the question of effective modelling this so called human uncertainty emerges naturally.

Our contribution introduces two different approaches for modelling the human uncertainty of user responses.
In doing so, we develop techniques in order to measure this uncertainty at the level of user inputs as well as the level of user cognition. With support of comprehensive user experiments and large-scale simulations, we systematically compare both methodologies along with their implications for personalisation approaches. Our findings demonstrate that significant amounts of users do submit something completely different (action) than they really have in mind (cognition).
Moreover, we demonstrate that statistically sound evidence with respect to algorithm assessment becomes quite hard to realise, especially when explicit rankings shall be built. 
\end{abstract}

%
%


\keywords{Human Uncertainty; Measurement Uncertainty; Identifiability; Distinguishability}
\maketitle

\newpage

\section{Introduction}

A broad range of algorithms and approaches in data mining aim at modelling and predicting aspects of human behaviour. 
These efforts are motivated by many practically relevant applications, including various recommender systems, content personalisation, targeted advertising, along with many others. This involves implicit or explicit knowledge about user behaviour, either by observing user interactions or by asking users explicitly.

We take this as an opportunity to ask the question of how people actually proceed when making decisions (e.g. creating ratings or other forms of feedback) while interacting with information systems. For example, many users may meet their decisions with considerable uncertainty in many situations, i.e. they would not exactly reproduce their decisions when asked twice or multiple times. This \textbf{Human Uncertainty}, as we understand it in this contribution, appears to be a characteristic feature of the cognitive process of decision making which influences its outcome, making it circumstantial and temporally unstable; the outcome appears to be more or less fluctuating randomly when repeating a decision making. Consequently, we may assume that observed decisions are drawn from individual distributions. Moreover, and even more important, our knowledge about such distributions may be very limited, due to natural limitations of known measurement methodologies.
One of these methodologies, which has already been used in recent research on data mining, is based on a frequentist approach and observes repeated user actions. Another approach, yet insufficiently discussed in this context, is based on a Bayesian approach and requires user perceptions. Both of these approaches have so far not been discussed sufficiently in the field of user modelling and personalisation. However, we will demonstrate that there are far-reaching implications of such considerations, especially for the statistical evidence of data mining results and the sometimes associated monetary decisions (e.g. when opting for the better recommender).

\begin{figure}[t]
    \centering
    \begin{subfigure}{0.35\textwidth}
        \includegraphics[width=\textwidth]{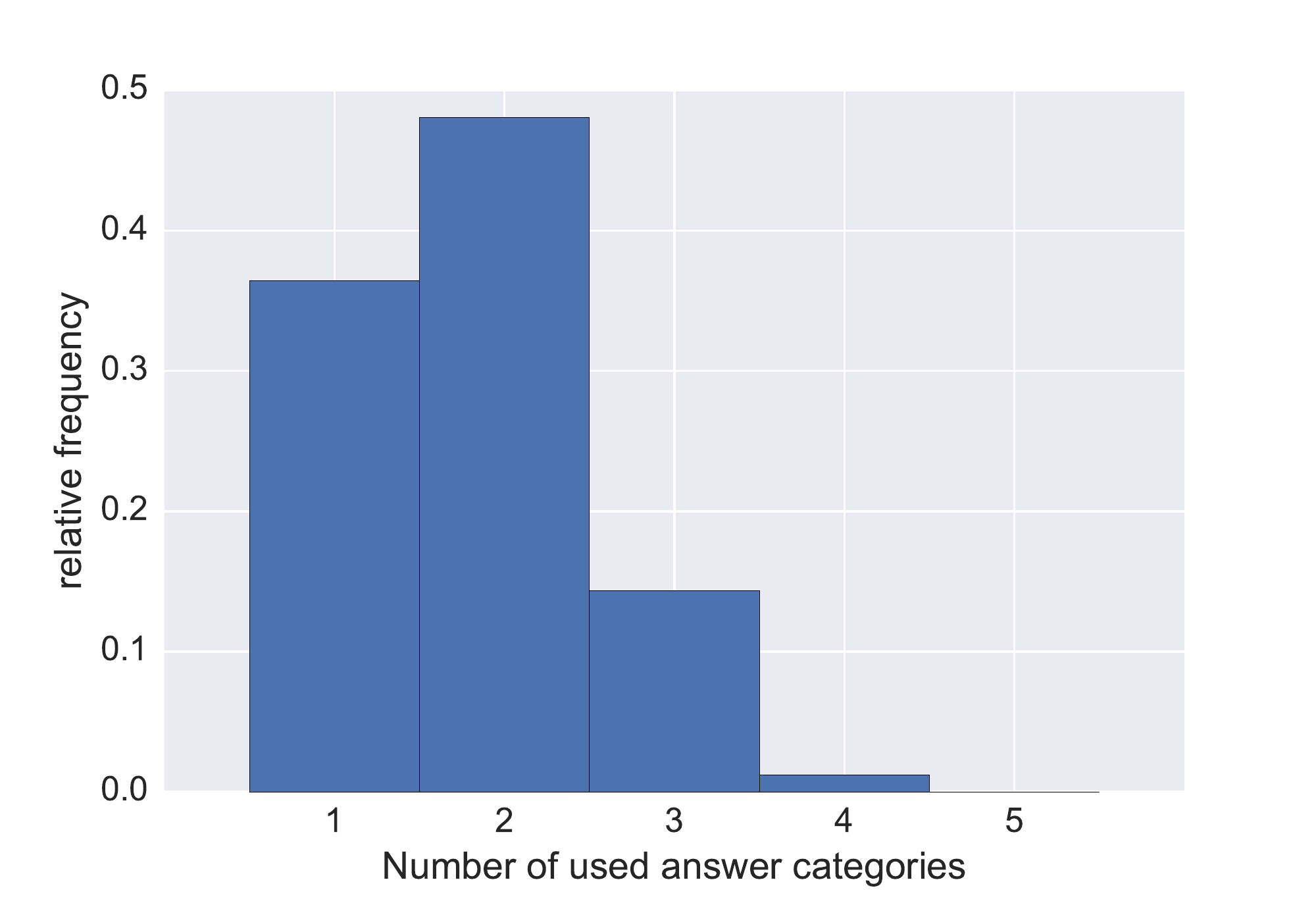}
        \caption{Frequency of the number of used answer categories}
        \label{fig:IntroA}
    \end{subfigure}
    \hfill
    \begin{subfigure}{0.35\textwidth}
        \includegraphics[width=\textwidth]{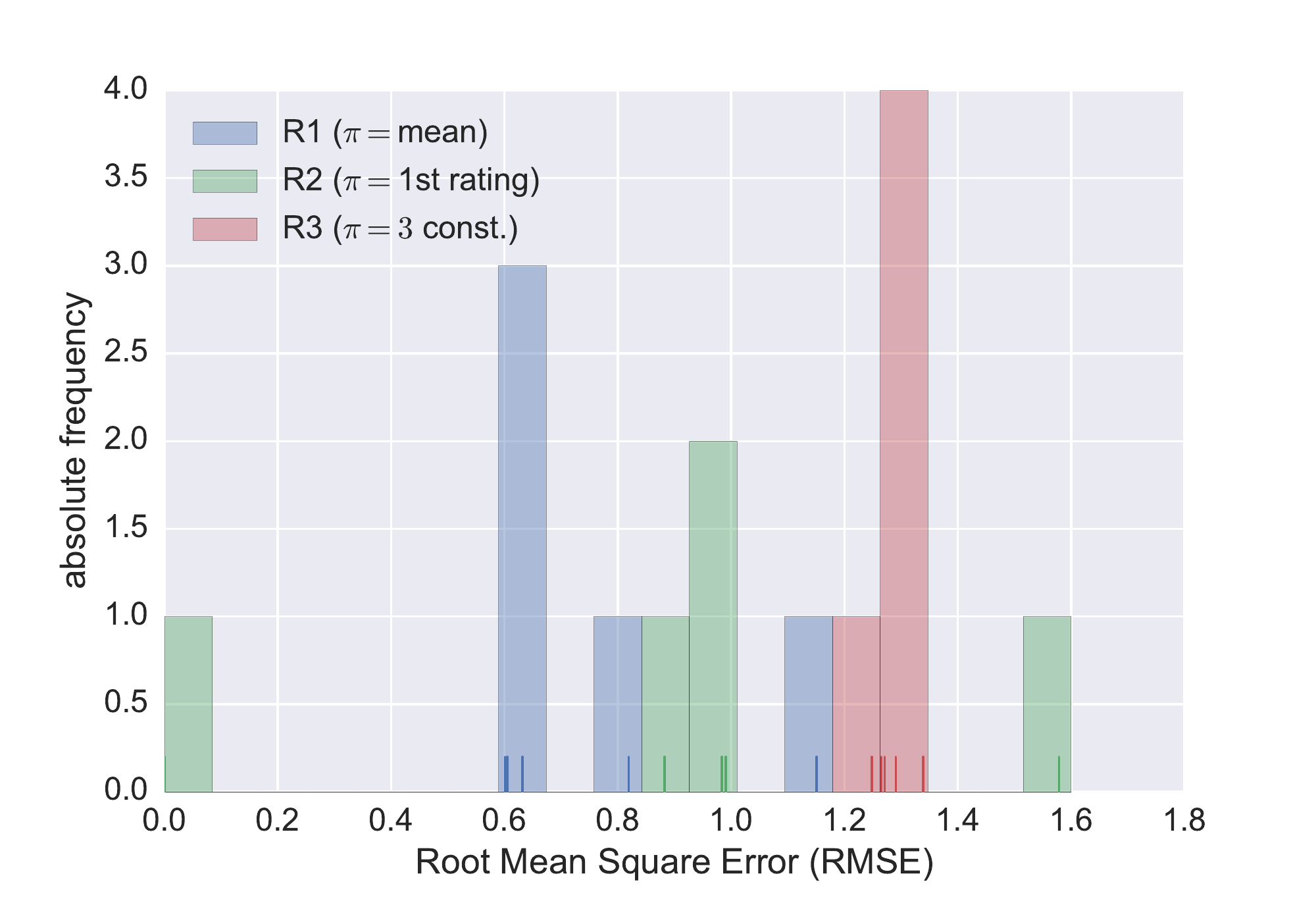}
        \caption{Distribution of RMSE outcomes for any rating trial}
         \label{fig:IntroB}
    \end{subfigure}
    \caption{Uncertain user ratings and impact on the RMSE}
\end{figure}

\paragraph*{Motivating Example} 
As a motivating example, we consider the task of rating prediction (common to recommender systems), along with the Root Mean Square Error (RMSE) as a widely used metric for prediction quality. 
In a systematic experiment with real users (described in more detail in forthcoming sections), individuals rated certain movie trailers multiple times. Figure \ref{fig:IntroA} shows that only 35\% of all users show constant rating behaviour, whereas about 50\% use two different answer categories and 15\% of all users make use of three or more categories. Based on these observations, we compute the RMSE for three recommender systems (designed by definition of their predictors $\pi$) for each rating trial. Figure \ref{fig:IntroB} depicts the RMSE outcomes and their frequency. It becomes apparent at once that the RMSE itself yields a particular degree of uncertainty, emerged from uncertain user feedback. When ranking these recommender systems, Figure \ref{fig:IntroB} allows for three possible results
\begin{equation}
(R1\prec R2\prec R3) \;\lor\; (R2\prec R1\prec R3) \;\lor\; (R1\prec R3\prec R2)
\end{equation}
depending on the rating trial, where the relation $\prec$ denotes  ``better than''. 
This problem is most obvious for recommender $R1$ as it could be both, the best or the worst recommender, although it operates for same users rating the same items. In addition, it may be possible that further repetitions of ratings would lead to even more ranking possibilities. This naturally implies to esteem the RMSE as a distribution rather than single scores. Consequently, the question changes from ``Is $R1$ better than $R2$?'' to ``How likely is it that $R1$ is better than $R2$?''.
Changing this question can be considered as a paradigm shift, i.e. from point-paradigm to distribution-paradigm, providing the possibility to detect and to visualise many interesting but so far hidden effects within the field of user modelling, personalisation and adaptation.

\paragraph*{The Problem}
Grounded on real user experiments it can be shown that there is often a significant overlap in two metric's distributions when assessing competing data mining approaches. During our analyses, we often encountered ranking error probabilities of 33\% or even more when evaluating by scores only. At this point, we must emphasise that this problem is not immanent to this novel paradigm, but has always been present implicitly in data records, provided that these are based on human behaviour. In fact, the distribution-paradigm is able to make some fundamental problems visible for the first time.
On these findings, it becomes imperative to explore possible propagations of human uncertainty in order to maintain  statistically sound and adequate methods of data mining.
However, in addition to the analysis of human uncertainty itself, the question of information quality in terms of reliability and validity also plays a major role and must be considered as well. As we're going to show in forthcoming sections, one has to repeat a rating task on the same item for a several hundred times in order to quantify the human uncertainty with acceptable precision.

To restate our problems in short:
We have to include human uncertainty in our decisions on algorithm assessment to exclude impacting errors, but with the most naive and simplest method to perform, we cannot gather enough information to quantify this uncertainty with necessary precision.

\paragraph*{Research Questions} 
In this contribution, we introduce two diametrically opposed approaches of gathering uncertainty information.
Due to the lack of sufficiently profound discussions in the literature of computer science that addresses human uncertainty, the compatibility of cognition and action, and the impact of these topics on commonly accepted techniques in data mining, we want to examine the relevance of this subjects in more detail.
In this spirit, we will focus on the following research questions:
\begin{description}
\item[Research Question Q1:] How do actual feedback responses differ from intended ones (in terms of probabilistic approaches)?\\[.1ex]
\item[Research Question Q2:] What implications become apparent by contrasting diverse uncertainty models (e.g. the impact on evaluation metrics like the RMSE)?
\end{description}
These questions are examined on the basis of user experiments that mimic the task of recommender systems.
The indications and implications presented in this contribution are nevertheless not limited to this field but do apply for most of common data mining approaches that explicitly account for human feedback.

\section{Related Work}
In this paper, we exemplify our approach using recommender systems \cite{handbook} and focus specifically on the validity of human uncertainty measurements in rating scenarios. 

The relevance of our contribution arises from the fact that the ubiquitous human uncertainty sometimes has a vast  influence on the evaluation of different prediction algorithms. For this comparative assessment, different metrics are used to determine the prediction quality, such as the root mean squared error (RMSE), the mean absolute error (MAE), the mean average precision (MAP) and many others \cite{workshop12}. These and other quality-related quantities in recommender assessment (e.g. user satisfaction, precision/recall, etc.) are summarised in \cite{Herlocker}. 
Although we exemplify the impact of measurement validity on the RMSE, the main results of this contribution can be easily adopted for alternative assessment metrics without substantial loss of generality, since they all share in common the need for uncertain human input.

The idea of uncertainty is not only related to predictive data mining but also to measuring sciences such as metrology. Recently, a paradigm shift was initiated on the basis of a so far incomplete theory of error \cite{Grabe}.
In consequence, measured properties are currently modelled by probability density functions and quantities calculated therefrom are now assigned a distribution by means of a convolution of these densities. This model is described in \cite{GUM}. A feasible framework for computing these convolutions via Monte-Carlo-simulation is given by \cite{GUMsupp1}. We take this model as a basis for our own modelling of uncertainty for addressing similar issues in the field of computer science.

The complexity of human perception and cognition can be addressed by means of latent distributions. This idea is widely used in cognitive science and in statistical models for ordinal data. For example, so-called CUB models for ordinal data \cite{cub} assume the Gaussian as a latent response model underlying the observations. We adopt the idea of modelling user uncertainty by means of individual Gaussians following the argumentation in \cite{cub} for constructing our individual response models. 

The impacts of human uncertainty for recommendation results have been frequently discussed in recent work from different perspectives. Observations presented in \cite{noise1, noise2} have shown that it can significantly influence results of recommender evaluation. The methodology applied there is based on repeating rating scenarios for same users-items-pairs and represents the current standard in latest research such as \cite{RateAgain}.
In this paper, the same methodology is explored and compared to a new Bayesian method, which we have derived from the from latest research on cognitions of uncertainty in educational scenarios \cite{HeinickeJasberg}.

\section{Data Modelling} 

\subsection{The Re-Rating Proceeding}

One way of deriving a user's rating-distribution is based on the \textbf{frequentist definition of probability}, i.e. the probability of an event to occur is equal to its relative frequency for infinite trials. Deduced from this definition we receive a probability density function (PDF) by simply asking users to re-rate the same item several times and computing Maximum-Likelihood-Parameters for a given data model. We will refer to this scenario as the \textbf{re-rating-proceeding}.

In mathematical terms: Let $U\subseteq \mathbb{N}$ be a finite set of Users and $I\subseteq \mathbb{N}$ a finite set of items to be rated. Let $S=\{1,2,3,4,5\}$ denote the set of possible ratings on the commonly used five-star ordinal scale, then the tensor $r_{u,i,t}\in S$ represents the  $t^{\textit{th}}$ rating from user $u\in U$ for item $i\in I$ where $t=1,\ldots ,N$. By forcing user $u$ to rate the same item $i$ multiple times, we obtain the sample
\begin{equation}
	r_{u,i,\bullet} := \{ r_{u,i,t} \vert\, t=1,\ldots ,N \}
\end{equation}
representing $t$ draws from the random variable $R_{u,i}$.
The corresponding \textbf{rating-distribution} represented by the PDF $f_R\colon \hat{S}\to\mathbb{R}$ can be generated by performing the ML-algorithm for a chosen data model (e.g. Gaussians, CUB-Models, etc.), assuming a continuous scale 
$\hat{S}$ as well as a non-vanishing variance of $r_{u,i,\bullet}$.
We then denote the standard deviation $\sigma_{u,i}:= \sqrt{\operatorname{Var}(R_{u,i})}$
as the operationalised \textbf{human uncertainty} of user $u$ on item $i$ where $\mu_{u,i}$ is the location-parameter. 
For this uncertainty have to regard two facts:
\begin{itemize}
\item A single user rates multiple items with unequal precision and thus produces a user-specific distribution 
with draws $\{  \sigma_{u,i} \vert\, i \in I \}$ which we call the \textbf{user-specific noise} $\Sigma(u)$.
\item A single item is rated by multiple users having unequal precision, emerging an item-specific distribution 
with draws $\{  \sigma_{u,i} \vert\, u \in U \}$ which we call the \textbf{item-specific noise} $\Sigma(i)$.
\end{itemize}
From this point of view, the human uncertainty for a specific user and item can be seen as a realisation of the joint PDF of $\Sigma(u)$ and $\Sigma(i)$. 
The biggest advantage of the re-rating-proceeding is that the users can, on the one hand, stick to the usual procedure but repeat several times. This procedure therefore seems to be very feasible, although it might be assumed that repetitions of a certain rating task are limited. However, the data obtained is easy to process. 
The disadvantages arise when we leave the view of probability theory and take the view of statistics, for then we are not able to know the parameters accurately, since we only calculate them on samples rather than the population. As a result, the parameters itself are subject to a so-called measurement uncertainty. In other words, we cannot  measure the human precision in sufficient quality, but only locate it within confidence intervals. This measurement uncertainty becomes an important factor since it propagates in every quantity derived from these rating-distributions. 

\subsection{The PDF-Rating Proceeding}
An alternative approach of accessing human precision is based on the Bayesian definition of probability, i.e.
the probability of an event to occur is the degree of one's personal confidence in this occurrence. 
Under this assumption, one can obtain a rating-distribution directly from requiring a user's personal confidence of the appropriateness for each possible rating that a scale provides. We will denote this procedure as the \textbf{pdf-rating-proceeding}. 

In mathematical terms: Having a 5-Star-Scale $S=\{1,\ldots,5\}$, a user associates to each possible rating $s\in S$ a degree of personal confidence about the appropriateness of $s$ concerning the item to be rated. The personal confidence is entered by a second Scale $S_C=\{1,\ldots,5\}$. Hence, a given rating 
\begin{equation}
r_{u,i}=\{(1,n_1),(2,n_2),(3,n_3),(4,n_4),(5,n_5)\}
\end{equation}
is given by a family of two-dimensional vectors in $S_L\times S_C$ where the values for ones personal belief are considered as specific weights for each of the associated ratings. In order to retrieve the rating-distribution, this rating is converted into its frequentist equivalent by use of the transformation 
\begin{equation}
\tau\colon r_{u,i} \mapsto (	\underbrace{1,\ldots,1}_{n_1\textrm{-times}}  , 
							\underbrace{2,\ldots,2}_{n_2\textrm{-times}} , \ldots ,
							\underbrace{5,\ldots,5}_{n_5\textrm{-times}} ).
\label{eq:MakeFreq}
\end{equation}
since the absolute histogram of this frequentist translation will exactly reproduce the data initially entered by the user.
We then perform a ML-Estimation on $\tau(r_{u,i})$ to find the optimal parameters for a chosen data model. The great advantage of the pdf-rating is, that all necessary data can be obtained by one rating only which grants a better viability and saves valuable time for improving the system, i.e. the machine-learning-process speeds up significantly. The disadvantages might be that this new and yet unusual method is not immediately accepted by users.


\subsection{Composed Quantities}
The RMSE, as a metric for model-based prediction quality, is a suitable example to demonstrate the impact of human uncertainty as well as the limited precision of its measurement on composed quantities. 
Composed Quantities, in this contribution, are quantities that compute as a function of large amounts of uncertain arguments. So, a composed quantity becomes a random variable itself whose density function emerges as a convolution of the density functions of its arguments. 

For further considerations, we assume all ratings to be normally distributed random variables 
$R_{u,i}\sim\mathcal{N}(\mu_{u,i},\sigma_{u,i})$ (rationally, it exhibits maximum entropy along all distributions with finite mean/variance and support on $\mathbb{R}$). 
Accordingly, the RMSE  
\begin{equation} \label{eq:RMSE}
\text{RMSE} = \sqrt{\sum_{(u,i)\,\in\,\mathcal{U}\times\mathcal{I}} \frac{(\pi_{u,i}  - R_{u,i} )^2}{n}}
\end{equation}
materialises as a composition of continuous maps of random variables and thus becomes a random variable itself. 
Its distribution emerges as a convolution of $n\leq U\cdot I$ density functions and computations can be easily done via Monte-Carlo-Simulation \citep{GUMsupp1}. 

In case of exactly known rating-distributions, we get a clear distribution for the RMSE.
However, since each dataset represents only one sample rather than the entire population, point estimators are inappropriate here. Instead, confidence intervals have to be specified \cite{VarConfInt}.
In that sense, we cannot simply determine a single rating-distribution for each user-item-pair, but have to compute a variety of distributions with the associated parameters drawn from corresponding confidence intervals. 
In consequence, even for large-scale computations, the resulting RMSE does not possess a stable density function. However, there exist borderline cases which reveal the maximum range in which we can expect results for the density function of the RMSE.

\section{Methodology and Experiments}

\subsection{The Experiment}
Our experiment is set up with Unipark's\footnote{http://www.unipark.com/de/} survey engine whilst our
participants were committed by the crowdsourcing platform Clickworker\footnote{https://www.clickworker.de/}.
During the experiment, participants watched theatrical trailers of popular movies and television shows and provided ratings 
using the re-rating-proceeding and pdf-ratings-proceeding respectively\footnote{A full description can be found on https://jasbergk.wixsite.com/research}. The submitted ratings have been recorded for five out of ten fixed trailers so that the remaining trailers act as distractors triggering the misinformation effect, i.e. memory is becoming less accurate because of interference from post-event information. 


\subsection{Evaluation Methodology}

\paragraph*{Research Question Q1:}

Here, we examine the difference between actual and intended ratings as obtained by the re-rating and pdf-rating respectively. To this end, we compare the rating distributions as well as the distributions of the variances (userspecific and itemspecific noise) resulting from the different measurement methods.

\vspace{1.25ex}

\textbf{Equality of distributions:}
To compare the distributions induced by actions as well as cognitions, we consider point-estimation parameters and go on three factors: On the one hand, we compare the distribution type by means of a two-sample-KS-test.
Even if the equality of two PDFs has to be rejected, the available rating-distributions may nevertheless possess the same expectation or variance that could be assigned to a user within a recommender system for a future rating. Therefore we perform Welch's t-test to compare the expected values as well as Levene's test to investigate homoscedasticity. It will turn out that all item-specific-noise distributions retrieved from the PDF-rating-procedure share a conspicuous common feature: The equality of expectations throughout all items. This hypothesis is explored by Welch's t-test as well. All testing is performed with a significance level of $\alpha=0.05$.

\vspace{1ex}

\textbf{Validity of distributions:}
We will also focus on validity of the distributions, that is the precision with which the relevant parameters can be localised. For all rating-distributions this can be done easily by comparing the length of the parameters' confidence intervals, due to an explicitly given parametric data-model. For the Noise-Distributions we do not have these parametric data-models. Instead, a Monte-Carlo-Simulation is used in which we sample the variances from their confidence intervals. For every resulting noise-distribution we then compute the percentiles $q\in[0,1]$, so that re-sampling will result into a distribution for each of this percentiles. In doing so, the standard deviation $\sigma(q)$  of $q$ becomes a measure for the precision with which the noise-distribution can be obtained. Thus, we simply compare the quantiles' standard deviations when deduced from the re-rating and pdf-rating.

\paragraph*{Research Question Q2}

Here, we examine implications of human uncertainty and their visualisation by a given measurement method.
For this purpose, we will focus on evaluation metrics and discuss the possible implications for the RMSE as an example. In particular, we will investigate the influence of measuring precision and the distinguishability of two RMSE-distributions.
Due to the fact that it is quite difficult to specify a closed form for the RMSE's density function \cite{Chan}, we will perform a Monte-Carlo-Simulation as described in \cite{GUMsupp1}. In our simulations, we observe six different recommender systems, designed by defining their predictors via 
\begin{equation}
\pi^k_{(u,i)}:=
\begin{cases}
\text{mean within all rating trials} & k=1 \\
 (k-1)^\text{th}\text{ rating}  & k=2,\ldots,6 
\end{cases}
\end{equation} 
where $k$ denotes the $k$-th recommender system.

\vspace{1.25ex}

\textbf{Equality of distributions:}
Since the MC-simulation is an artificial generation of draws, hypothesis testing can not be executed directly on this data set in order to validate whether both measurement methods produce the same RMSE or not. This is because a simple increase in the MC-trials can be used to significantly detect any effect, even if this is not possible from the underlying data set. For this reason, we freeze the parameters of the incoming variables to the corresponding point estimators and indirectly simulate the hypothesis tests. To this end, we reduce the number of MC-trials to the actual sample size of the collected data and carry out the hypothesis test on these reduced samples. However, we repeat this $10^6$ times and consider the relative frequency $h$ of the rejection of equality. If $h>\alpha-1$ holds, a possible effect can be considered to be proven with significance level $\alpha$.

\vspace{1ex}

\textbf{Validity/Reliability of distributions:}
In case of the RMSE, validity is closely linked to reliability as the inaccurate location of the rating-distributions' parameters (validity) induces diverse outcomes for any re-sampling (reliability). The validity in terms of precision is observed by sampling the percentile's distributions again and compare their standard deviations when sampled from the re-rating-proceeding as well as the pdf-rating-proceeding. The effect size of reliability will be demonstrated by considering borderline cases of the RMSE which reveal the maximum span in which we can expect results for its density function.

\vspace{1ex}

\textbf{Distinguishability:}
Our analyses will reveal that two recommender systems $R1$ and $R2$ may not only have different PDFs $f_{R1}(x)$ and $f_{R2}(x)$ for the RMSE, but also do these PDFs overlap very often.
Thus each ranking $R1\prec R2$ is always subject to an error
\begin{equation}\label{eq:ErrorProb}
P_\varepsilon :=P(R1>R2) = \int_{-\infty}^\infty f_{R2}(x) \big( 1-F_{R1}(x) \big) \,\mathrm{d}x 
\end{equation}
where $F_{R1}(x)$ denotes the cumulative distribution function of the RMSE-distribution of $R1$.
In this context, we investigate how strongly one recommender must deviate from another, so that this can be significantly recognised by the RMSE, i.e. the probability of ranking error diminishes to less than five percent. To this end, we assume to have perfectly known rating-distributions and compute the RMSE for two recommender systems with adjustable prediction quality. Theoretically, the arithmetic mean $\bar{x}_{u,i}$ of ratings throughout all rating trials appears to be the optimal predictor, because this is the value which is obtained on average in the case of an infinite repetition and thus produces the smallest sum of squared deviations. Hence, we define the optimal recommender by setting $\pi_{u,i}:=\bar{x}_{u,i}$. To this optimum we additionally create a copy which we distort by artificial noise generated from re-sampling its predictors $\pi_{\text{new}} \in [(1-p) \pi_{\text{old}} \,;\, (1 + p) \pi_{\text{old}}]$. In this case, a noise fraction of $p$ means that those new predictors deviate from the originals by $(100\cdot p)$\%. The RMSE thereby receives a shift on the x-axis so that it's possible to calculate a declining error probability $P_\varepsilon(p)$ for a given ranking. In this process, we observe the amount of noise that is necessary to fulfil the distinguishability-condition $P_\varepsilon < 0.05$.

\subsection{Results}
Altogether $67$ people from Germany, Austria and Switzerland participated in this experiment. This group can be parted into $57\%$ females and $43\%$ males whose ages range from $20$ to $60$ years whilst over $60\%$ of our participants where aged between $20$ and $40$. This group also includes a good average of lower, medium and higher educational levels. The rating frequency habits range from rarely to often in uniform distribution. According to this data we can assume to have gathered a cross-sectional data, generally reflecting the German speaking population from these three countries.


\paragraph*{Research Question Q1}
The comparison of rating-distributions deduced from actions and cognitions in terms of the KS-test shows that descriptive deviations are not significant in 207 of 301 cases ($\approx 98\%$). The comparison of expectations by means of Welch's t-test reveals that these do not differ significantly from one another in 175 of 211 cases ($\approx 83\%$). Similarly, Levene's test shows that a deviation from homoscedacity was only significant in 175 of 211 cases ($\approx 82\%$). A more detailed breakdown by items is given in Table \ref{tab:Equality1}. Overall, the probabilistic ratings may indeed possess descriptive deviations, but all are within the range of random fluctuations.
\begin{table}
\begin{small}
\renewcommand{\tabcolsep}{1mm}
\begin{tabular}{c|cc|cc|cc|}
\cline{2-7}
& \multicolumn{2}{c|}{KS-Test} & \multicolumn{2}{c|}{Welch's t-Test} & \multicolumn{2}{c|}{Levene-Test} \\ \cline{2-7}  & n. rejected 		& rejected 	& n. rejected 	& rejected 		& n. rejected 	& rejected \\ \hline
Item 1 	& 59 (1.00) 		& 0 (0.00) 	& 52 (0.88) 	& 7   (0.12) 	& 52 (0.88) 	& 7   (0.12)\\
Item 2 	& 39 (0.98) 		& 1 (0.02) 	& 34 (0.85) 	& 6   (0.15) 	& 33 (0.82) 	& 7   (0.17)\\
Item 3 	& 31 (0.94) 		& 2 (0.06) 	& 23 (0.70) 	& 10  (0.30)	& 26 (0.79) 	& 7   (0.21)\\
Item 4 	& 45 (1.00) 		& 0 (0.00) 	& 38 (0.84) 	& 7   (0.16) 	& 37 (0.82) 	& 8   (0.18)\\
Item 5 	& 33 (0.97)			& 1 (0.03) 	& 28 (0.82) 	& 6   (0.18) 	& 26 (0.76) 	& 8   (0.24)\\ 
\hline
\end{tabular} 
\end{small}
\caption{Hypothesis testing on the rating-distributions -- absolute counts first, fractions in brackets}
\label{tab:Equality1}
\end{table}

\begin{figure*}[t]
    \centering
    \begin{subfigure}{0.32\textwidth}
        \includegraphics[width=\textwidth]{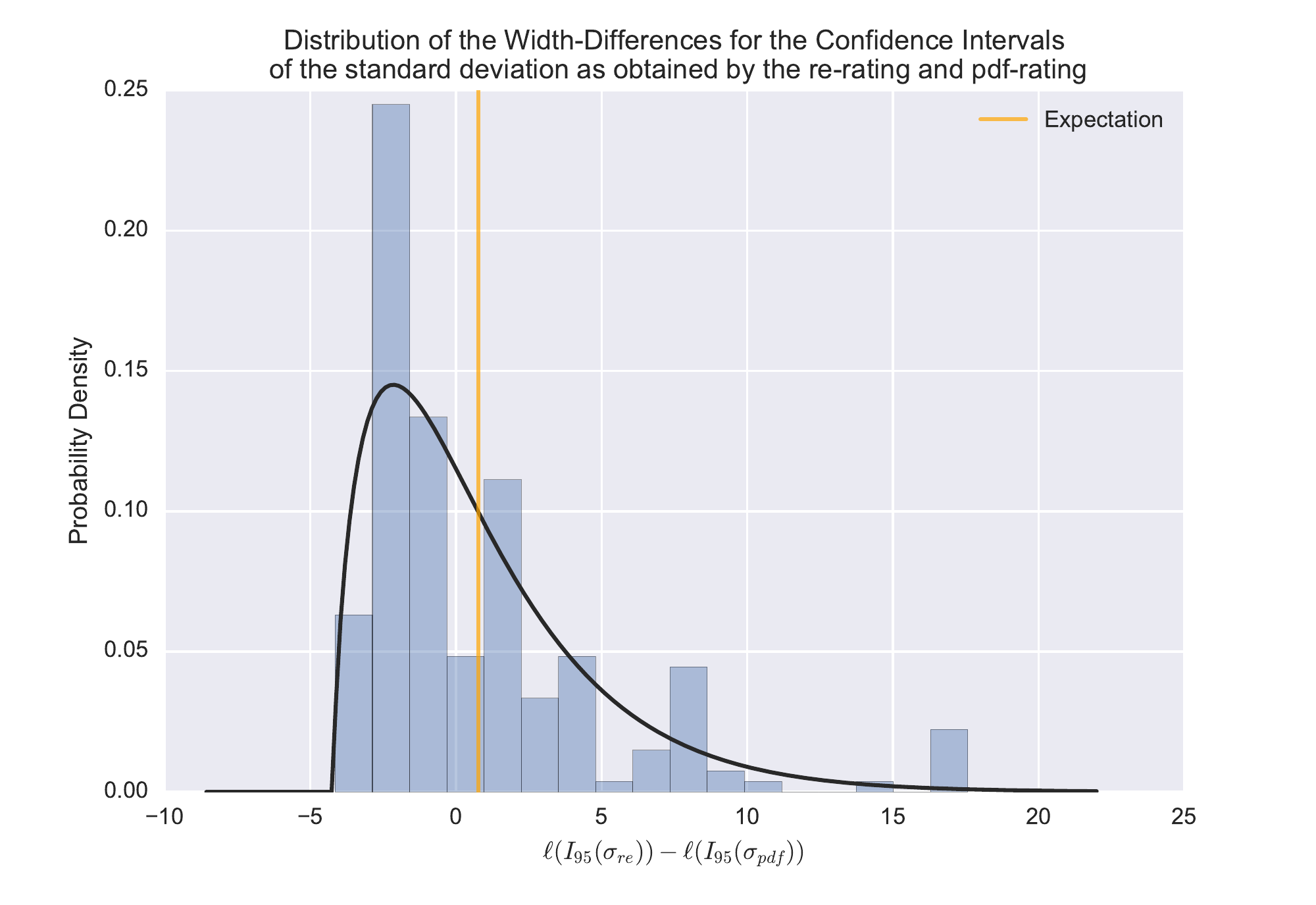}
        \caption{Distribution of length differences for the 95\%-intervals of the mean value}
    \end{subfigure}
    \hfill
    \begin{subfigure}{0.32\textwidth}
        \includegraphics[width=\textwidth]{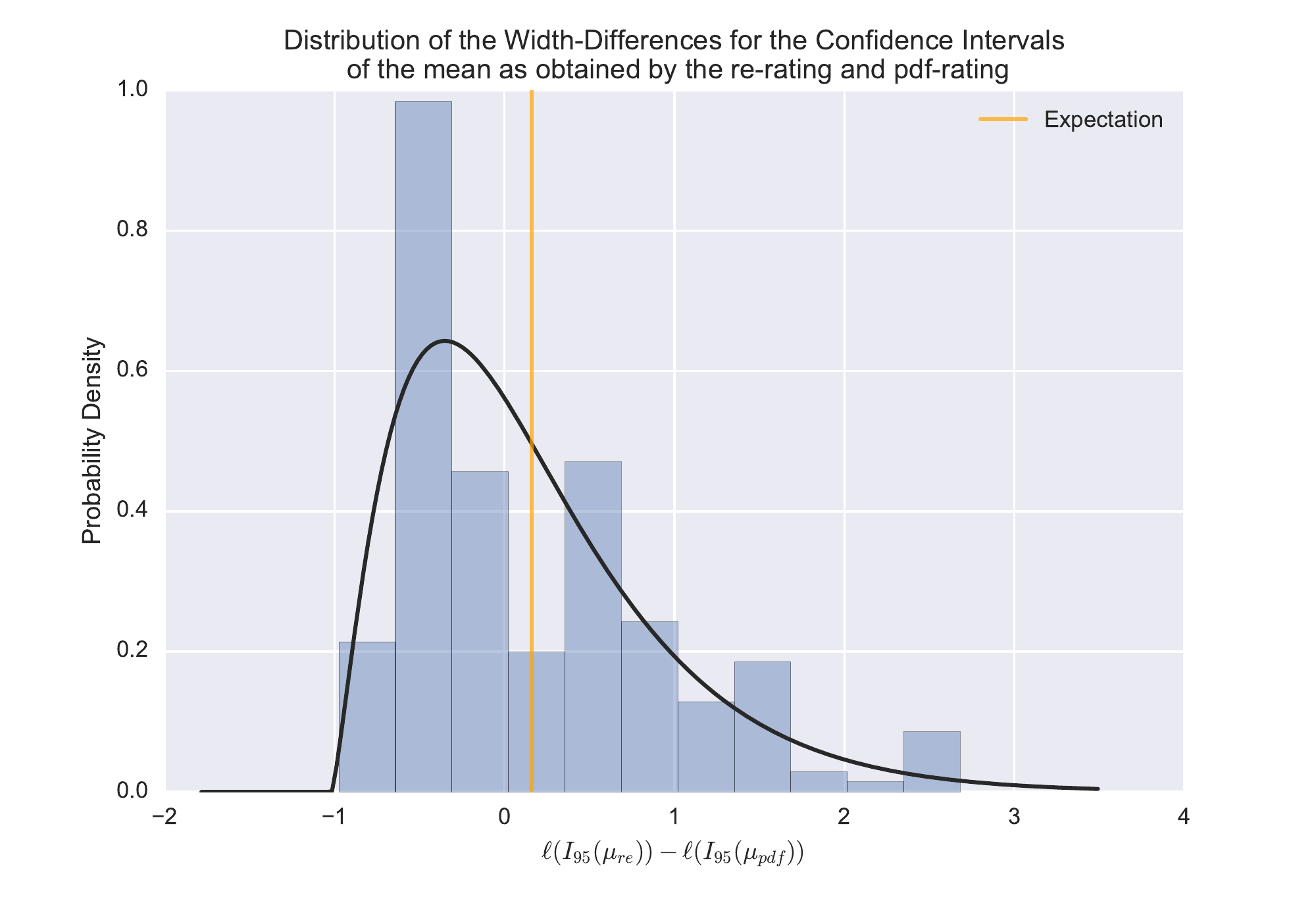}
        \caption{Distribution of length differences for the 95\%-intervals of the standard deviation}
    \end{subfigure}
    \hfill
    \begin{subfigure}{0.32\textwidth}
        \includegraphics[width=\textwidth]{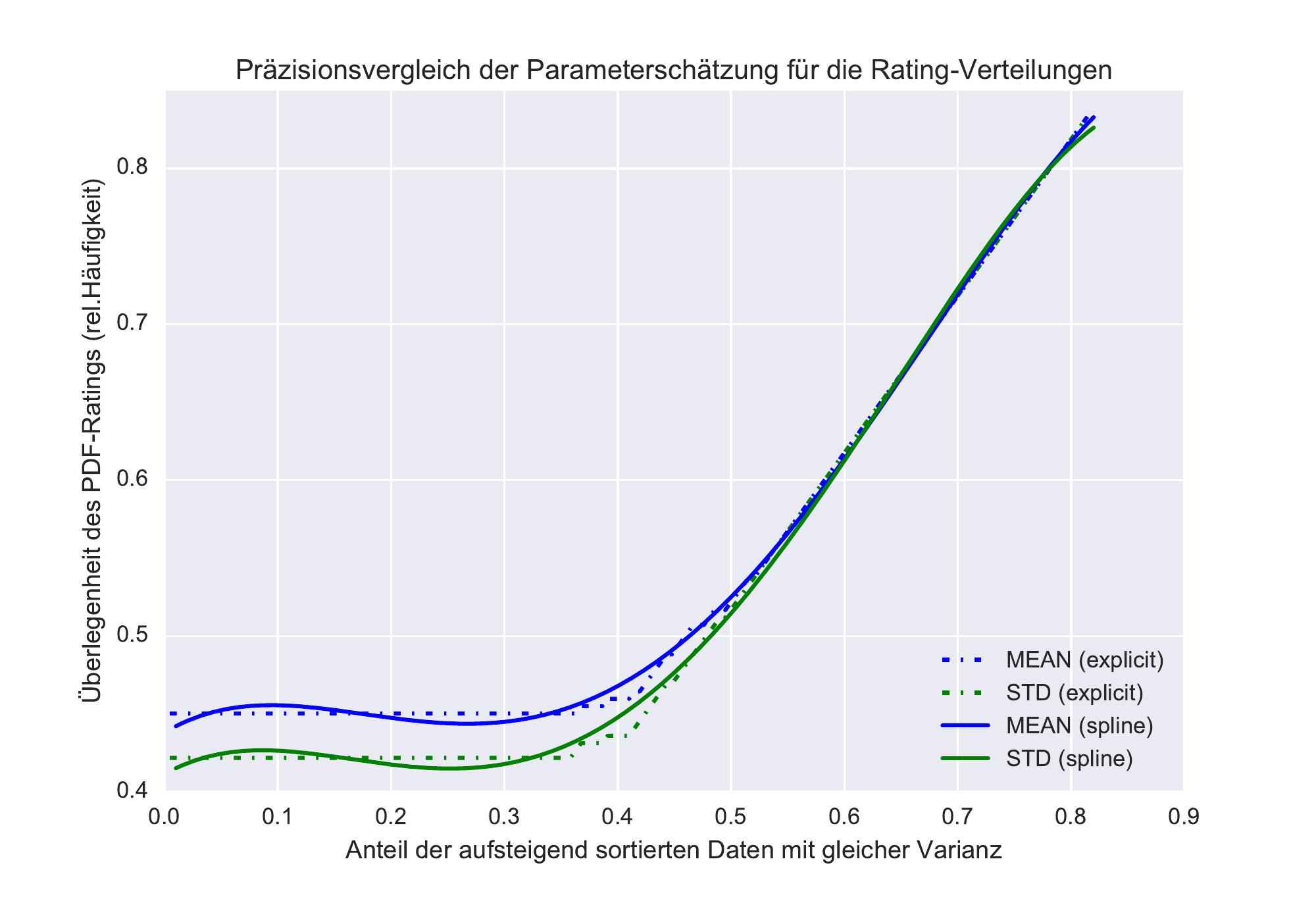}
        \caption{pdf-rating-superiority as a function of permitted equality of non-significant deviations}
    	\label{fig:RatingPrecisionC}
    \end{subfigure}
    \caption{Analysis of the rating-distributions' precision depending on the applied method of gathering uncertainty}
    \label{fig:RatingPrecision}
\end{figure*}

Unfortunately, any of the investigated distributions is ambiguous, since its parameters can only be located within confidence intervals. For the assessment of mean value precision, we consider this quantity for each rating distribution obtained from re-rating ($\mu_{r}$) as well as pdf-rating ($\mu_{p} $) together with the 95\%-intervals and compare their length with aid of the auxiliary variable 
\begin{equation}
\Delta_\mu:=\ell(I_{95}(\mu_{r})) - \ell(I_{95}(\mu_{p})).
\end{equation}
If $\Delta_\mu>0$, then the length $\ell(I_{95}(\mu_{r}))$ of the re-rating-interval is greater than the length $\ell(I_{95}(\mu_{p}))$ of the pdf-rating-interval, i.e. the pdf-rating appears to be more precise in locating the mean value. The analysis of the standard deviation is done analogously. Figure \ref{fig:RatingPrecision} depicts the distribution of these length differences. It can be seen that the mass-ratio of improvements and deteriorations is very balanced. At the same time, it can be seen that the strength of these deteriorations are small in comparison to the strength of the improvements. The expectations show that on the average, the pdf-rating will produce a slight increase of overall precision. However, this analysis is merely based on descriptive considerations so far. For both of the parameters $\mu$ and $\sigma$, hypothesis testing reveals that more than 80\% of all correspondences do not differ significantly. This certainly does not mean that equality can be accepted in all these cases, but this is a possibility that can't be rejected.
In Figure \ref{fig:RatingPrecisionC}, we gradually allow this possibility, beginning with the smallest differences. It turns out that the precision of the pdf-rating gains very quickly when the fraction of equalities for the non-significant deviations increases.

A comparison of noise distributions leads to a contrary result (see Table \ref{tab:Equality2}). For the item-specific noise as well as the user-specific noise, the corresponding distributions can be regarded as significantly different with respect to the different measurement methods.
\begin{table}
\begin{small}
\renewcommand{\tabcolsep}{1mm}
\begin{tabular}{c|cc|cc|cc|}
\cline{2-7}
& \multicolumn{2}{c|}{KS-Test} & \multicolumn{2}{c|}{Welch's t-Test} & \multicolumn{2}{c|}{Levene-Test} \\ \cline{2-7}  										& n. rejected 	& rejected 		& n. rejected 	& rejected 		& n. rejected 	& rejected \\ \hline
ISN  		& 0 (0.00) 		& 5 (1.00) 		& 0 (0.00) 		& 5   (1.00) 	& 1 (0.20) 	& 4   (0.80)\\
USN  	& 0 (0.00) 		& 67 (1.00) 	& 33 (0.49) 	& 34   (0.51) 	& 55 (0.82) 	& 12   (0.18)\\
\hline
\end{tabular} 
\end{small}
\caption{Hypothesis testing on the noise-distributions -- absolute counts first, fractions in brackets}
\label{tab:Equality2}
\end{table}
These findings are substantiated by the approximation of underlying data-models. 
In any case, the item-specific noise (ISN) is following power-law-distributions when measured via re-rating and respectively following Gaussians when measured via pdf-rating. These highly significant deviations indicate that the perceived noise differs form the observable noise. The power-law-distribution would claim that many people are quite certain whereas larger uncertainty only manifests for a few people. This doesn't perfectly match the everyday experience and may emerge due to the limitations of the conventional rating instrument, i.e. customers are forced to choose precisely one discrete rating option. But unfortunately this ``all or nothing'' does not match human cognitions when making decisions.
This interpretation is supported by the fact that an outstanding majority had chosen more than one answer category.
Thus by forcing a user to rate on discrete scales, the user will always select the mode of his ``inner distribution'' or perhaps some value nearby, depending on external influences. In contrast, the Gaussian data-model indicates, that many people possess the more or less the same uncertainty and that deviations in both directions are equally likely whereas large deviations are less likely than small deviations. In our experiments we located this common uncertainty to be $1.3$ stars. Welch's t-test indicates that only 10\% of all ISN distributions possess a significantly different expectation when using the pdf-proceeding. This is a strong indication for a latent cognition that is present for the majority of observed users.

\begin{figure*}[t]
    \centering
    \begin{subfigure}{0.3\textwidth}
        \includegraphics[width=\textwidth]{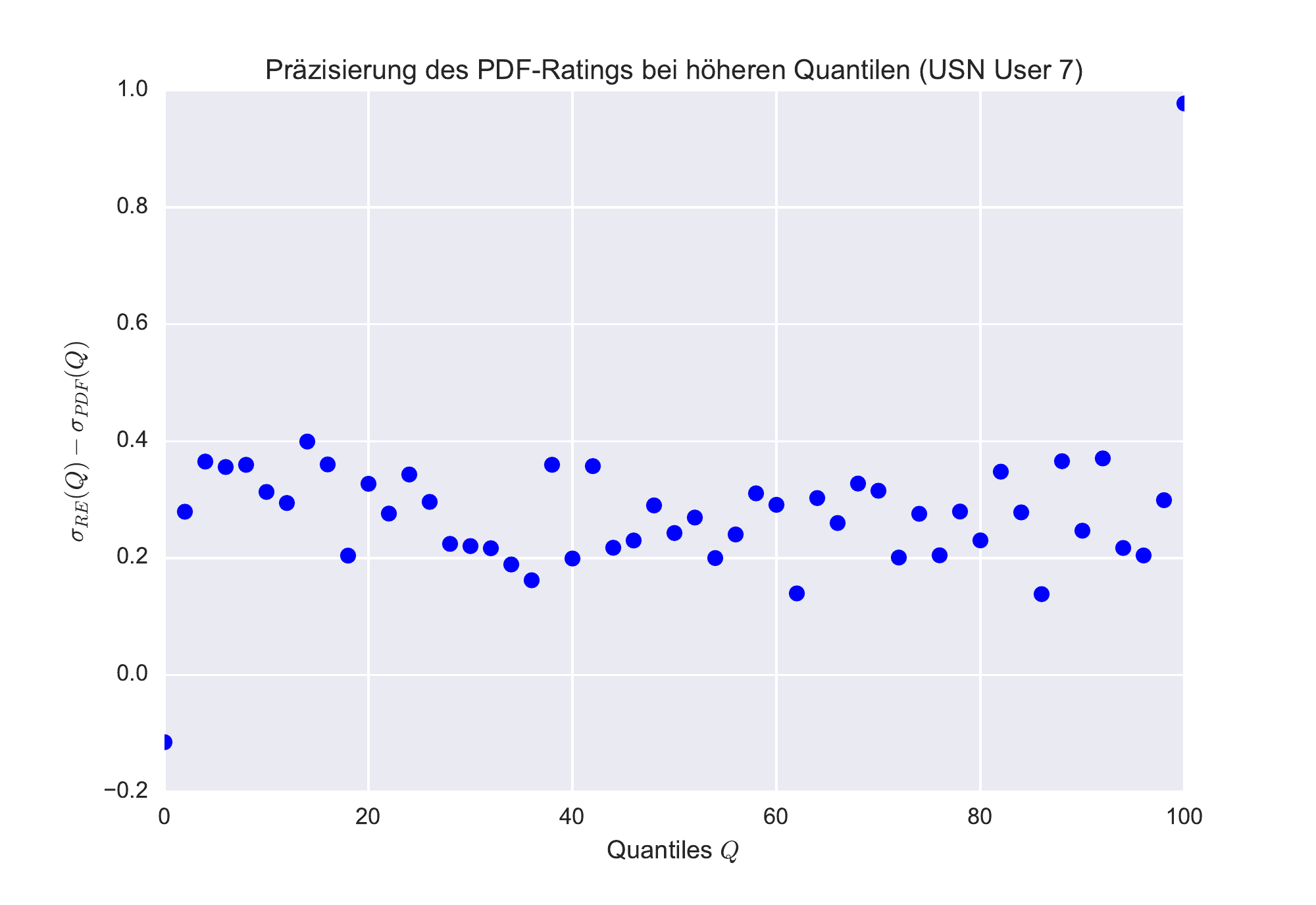}
        \caption{homogeneous}
        \label{fig:ArchetypesA}
    \end{subfigure}
    \hfill
    \begin{subfigure}{0.3\textwidth}
        \includegraphics[width=\textwidth]{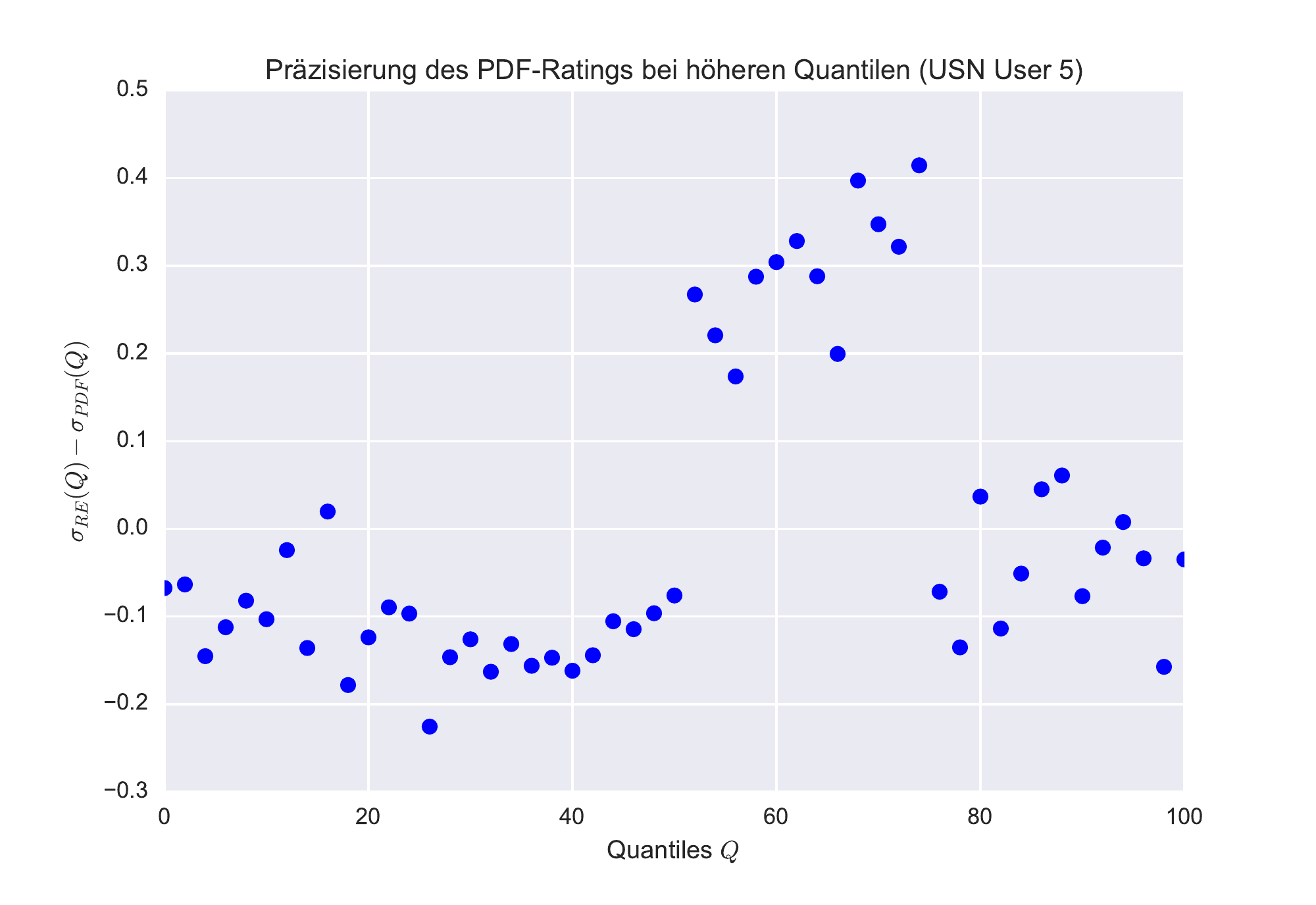}
        \caption{clustered}
        \label{fig:ArchetypesB}
    \end{subfigure}
    \hfill
    \begin{subfigure}{0.3\textwidth}
        \includegraphics[width=\textwidth]{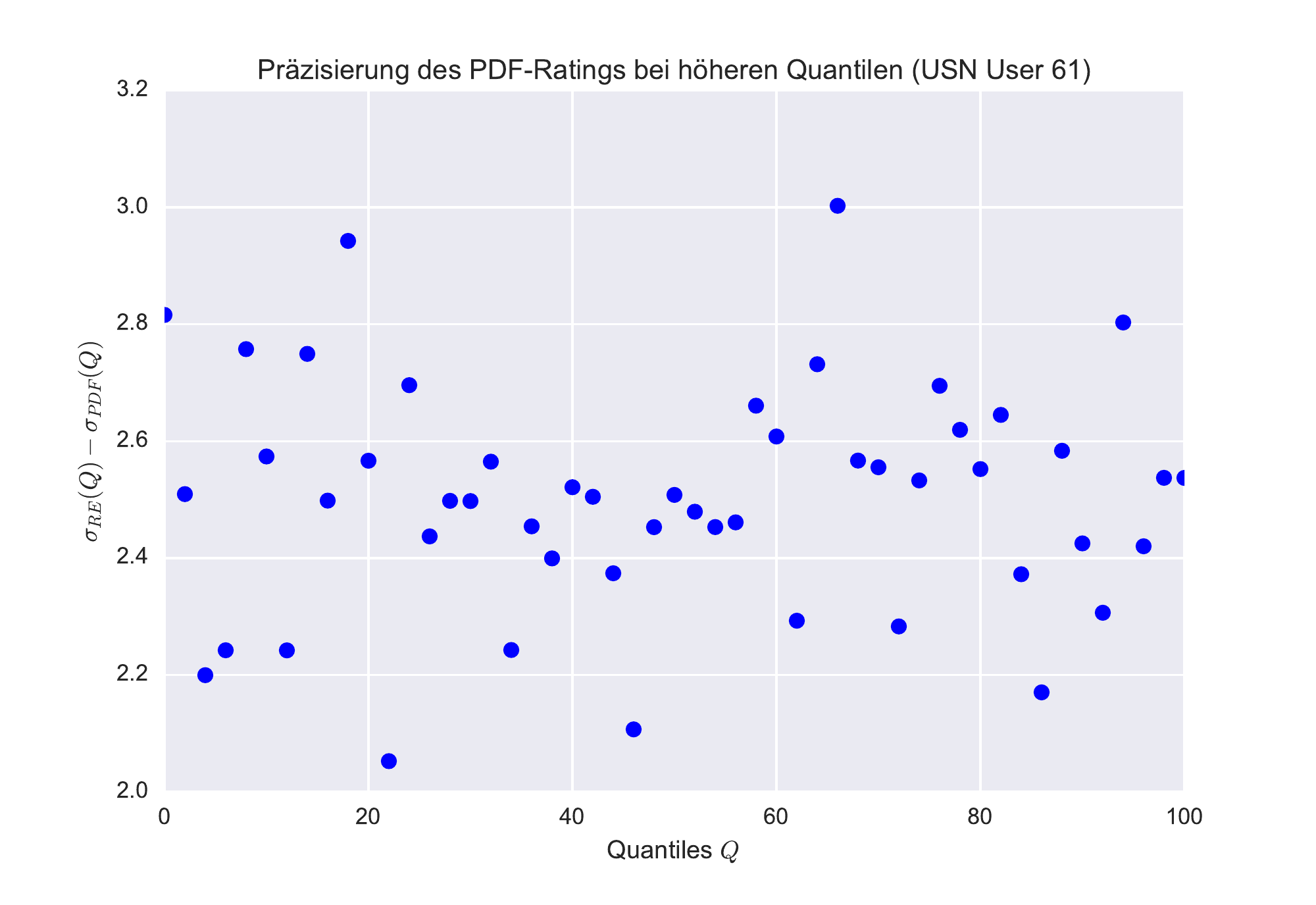}
        \caption{irregular}
    	\label{fig:ArchetypesC}
    \end{subfigure}
    \caption{Examples of user archetypes revealing ranges where cognition is more precise than action}
    \label{fig:Archetypes}
\end{figure*}

For our investigations of the validity in terms of measurement precision, we compute the percentile distributions for each corresponding ISN and USN by re-sampling. In doing so, the standard deviations of these percentile distributions naturally become a measure of noise precision. The auxiliary quantity $\delta_q:= \sigma_r(q)-\sigma_p(q)$ is positive for $\sigma_p(q) < \sigma_r(q)$ indicating superiority of the pdf-rating. A counting proves the invariance of the precision under measurement methodologies for lower percentiles as well as the superiority of the pdf-rating for higher percentiles.

The exploration of the USN validity must be considered more carefully. 
When computing scatter plots for $\delta_q$ against $q$, there are three always repeating archetypes to be spotted, which are monotonic behaviour (homogeneity), at least two clusters (clustered), and high dispersion with no visible relationship (irregularity).
Having these descriptions in mind, all scatter plots were independently assigned to a category by two analysts (inter-rater reliability $\varrho=0.99$). The quantitative extent of these archetypes within our data records summarises as follows: 28\% of all users can be considered homogeneous, 27\% can be considered irregular and 45\% of all users tend to be clustered. Representatives of these archetypes can be seen in Figure \ref{fig:Archetypes}. Homogeneous users (\ref{fig:ArchetypesA}) either show no significant effect (constant line) or a functional relationship, so that the uncertainty by action can be converted into uncertainty by cognition and vice versa.
Cognition and action are closely linked for these users, i.e. they make their decisions very thoughtfully and possibly not based on feelings. For the cluster archetype it can be seen, that those users allow for options in the pdf-rating which they would otherwise never have considered. Action and perception are not in harmony, i.e. these users probably use mainly their gut feeling for making decisions. The irregular archetype does not show any relationship between action and cognition.
Probably, those users have not rated seriously and just clicked through the online survey.

\paragraph*{Research Question Q2} 

Evaluation of the hypothesis testing indicates that the RMSE-distributions from both rating proceedings are fundamentally different for any of the simulated recommender systems. In particular, this is caused by a significant shift of the distributions' expectations, whereas the standard deviations does not differ significantly in any case.
Accordingly, it can be concluded that the measurement uncertainty mainly affects the location-parameter of the RMSE and that its spread can thus only be impacted significantly by the human uncertainty. 
This $\sigma$-invariance under measurement proceeding is consistent with our findings from research question Q1, i.e. that the rating-distributions (containing the human uncertainty information) are more or less equivalent. 

Nevertheless, the pdf-rating provides an information gain that might result into precision enhancement or reliability growth. When plotting the auxiliary variable $\delta_q$ against the sampled percentiles $q$ of the RMSE-distributions, the pdf-rating outperforms the re-rating in any case. This precision even increases monotonically for higher percentiles. Thus, the pdf-rating is theoretically supposed to limit the number of possible outcomes for any re-sampling of the RMSE-distribution. 
\begin{figure}[b]
    \centering
    \begin{subfigure}{0.23\textwidth}
        \includegraphics[width=\textwidth]{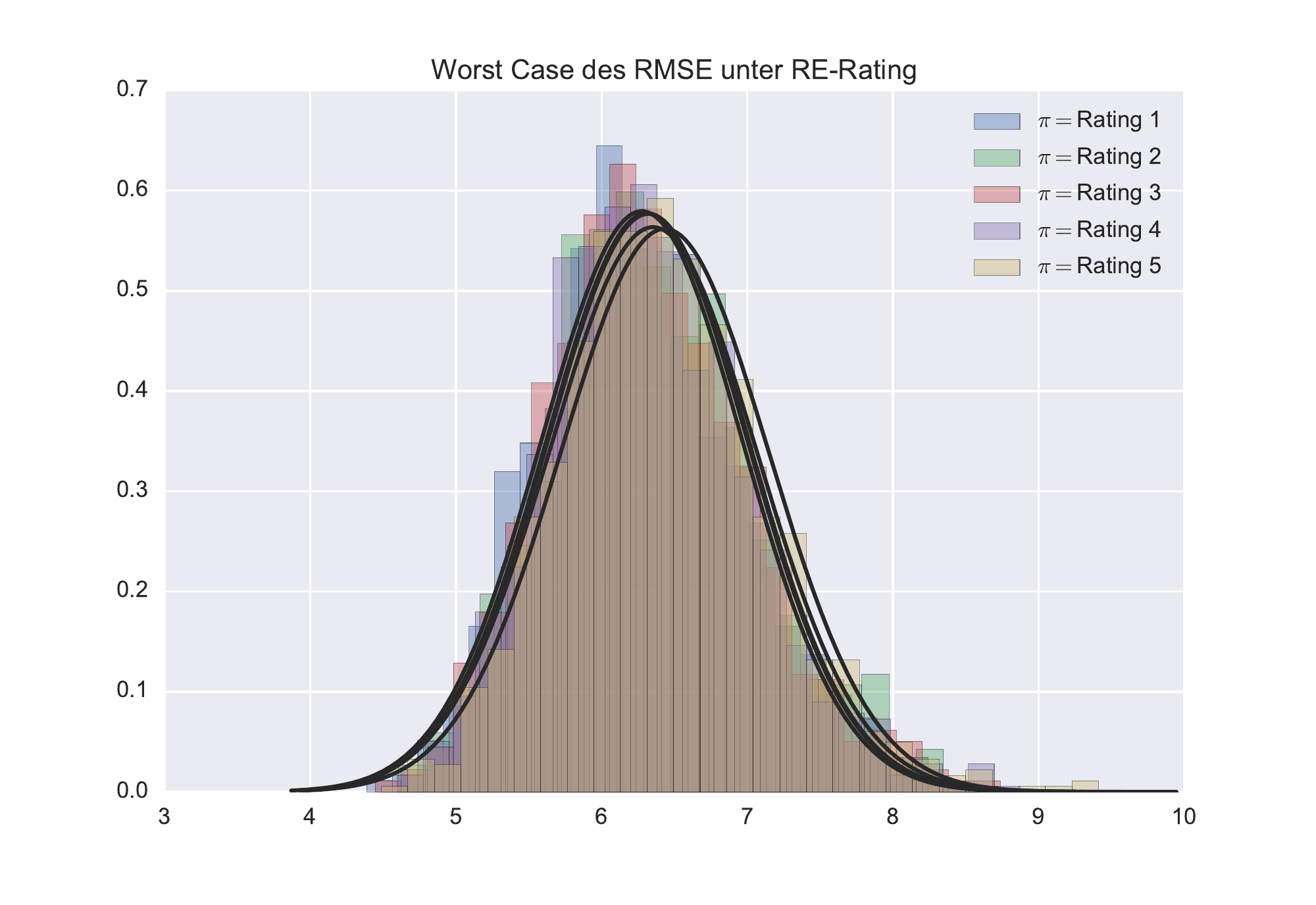}
        \caption{worst case for re-rating}
    \end{subfigure}
    \hfill
    \begin{subfigure}{0.23\textwidth}
        \includegraphics[width=\textwidth]{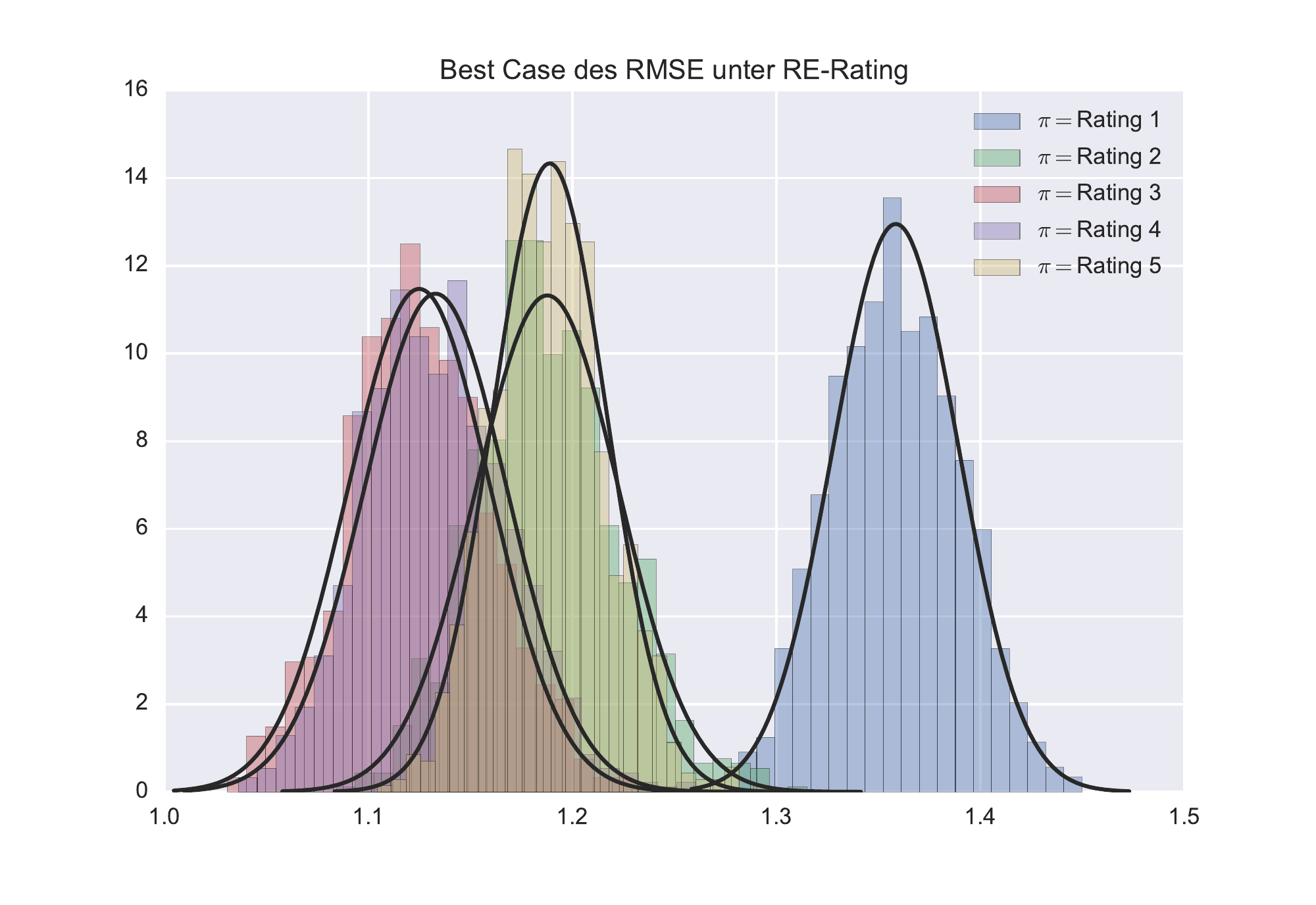}
        \caption{best case for re-rating}
    \end{subfigure}
    
    \begin{subfigure}{0.23\textwidth}
        \includegraphics[width=\textwidth]{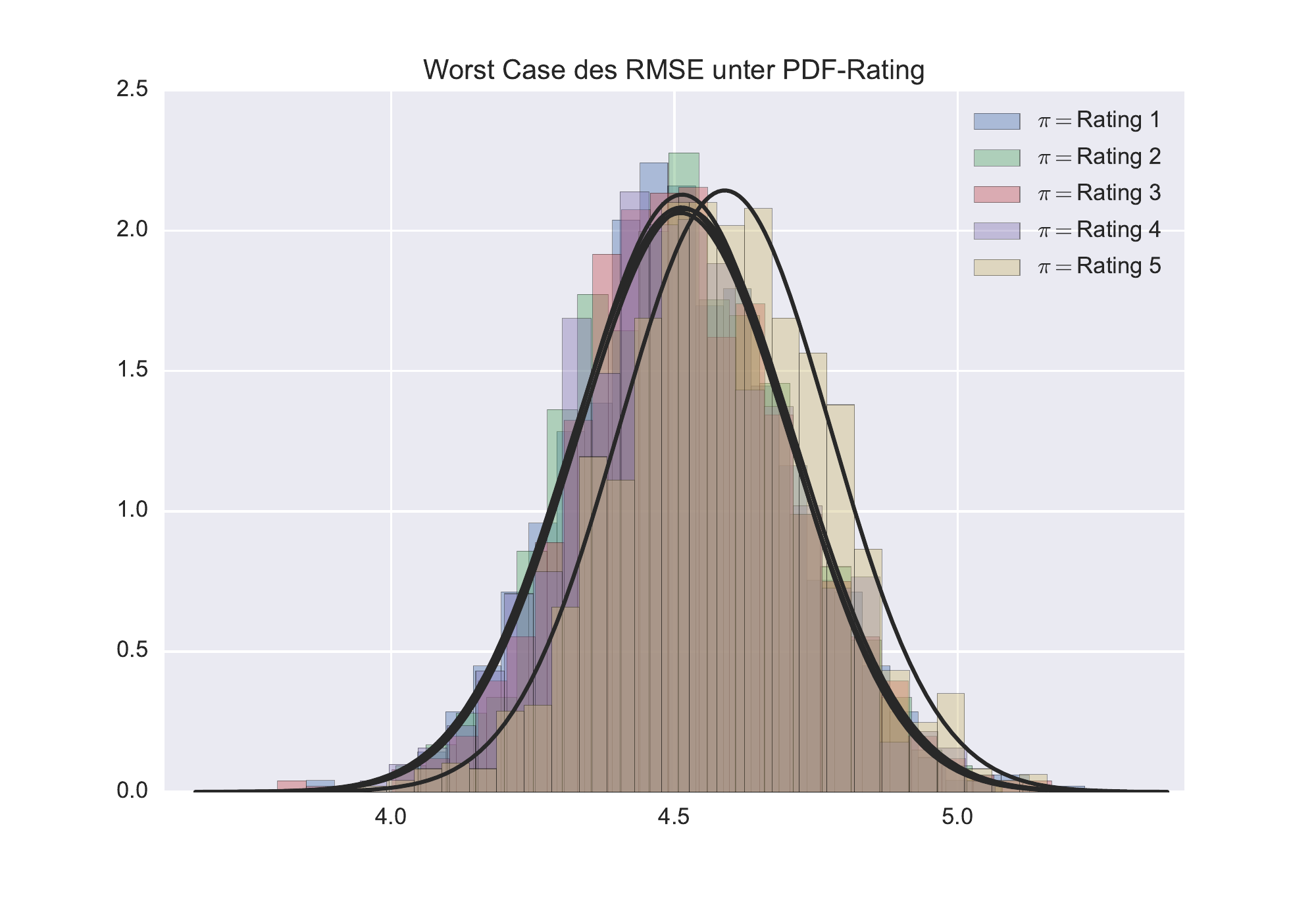}
        \caption{worst case for pdf-rating}
    \end{subfigure}
    \hfill
    \begin{subfigure}{0.23\textwidth}
        \includegraphics[width=\textwidth]{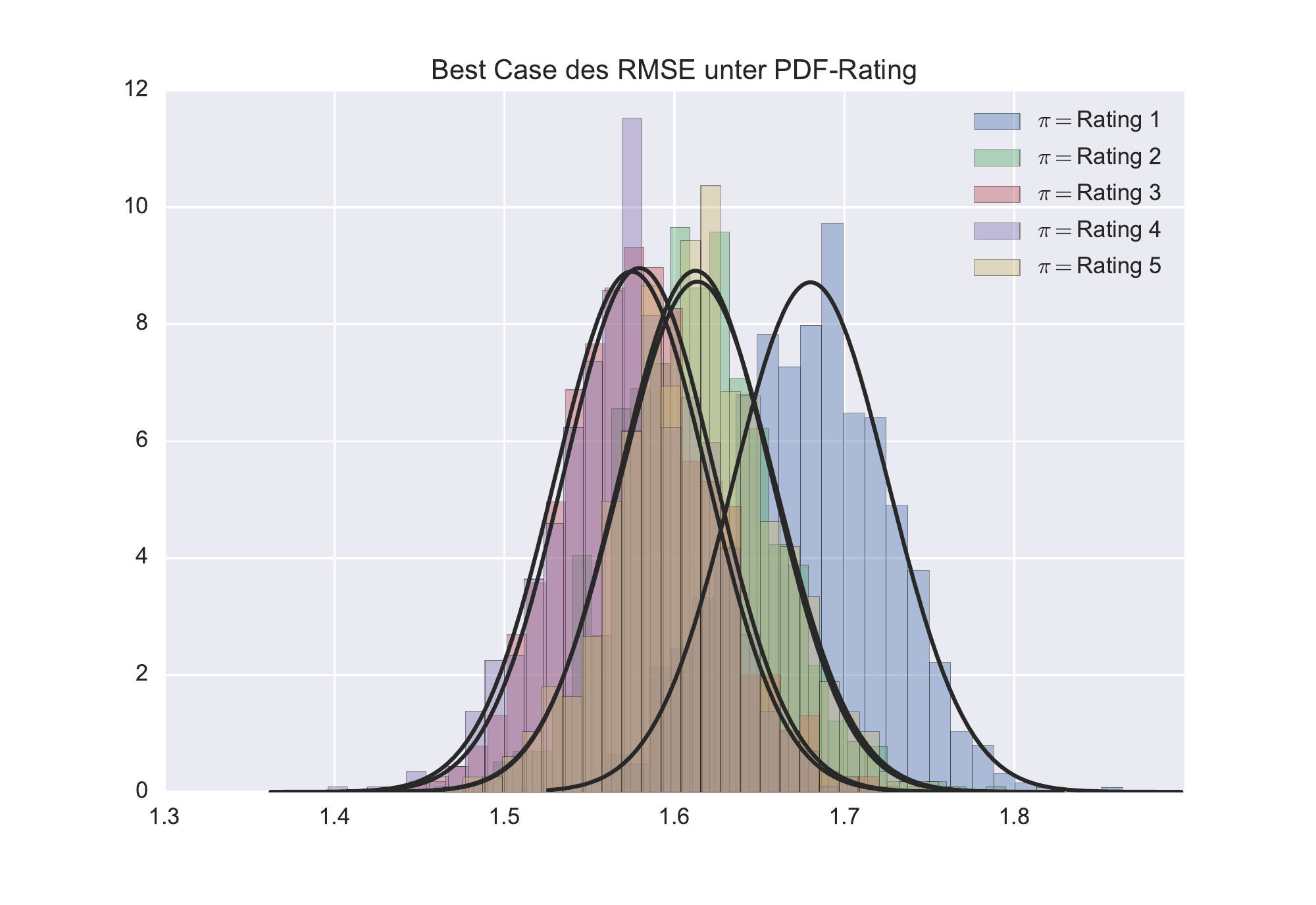}
        \caption{best case for pdf-rating}
    \end{subfigure}
    \caption{Borderline cases for the RSME of different recommender systems using re-rating and pdf-rating.}
    \label{fig:MinMaxRMSE}
\end{figure}

Figure \ref{fig:MinMaxRMSE} reveals the tremendous impact of both, human uncertainty and measurement uncertainty, on composed quantities. The ambiguity of the rating-distributions also lead to an ambiguity of the RMSE with a large range of possibilities. Whilst we can recognise a good resolution for a few RMSEs in the best case, this is virtually no longer possible for the worst case. The higher precision of the pdf-rating can also been observed here: Within the worst case, a density function is shifting to the right, so that the error probability decreases. Within the best case, the densities are moving closer to the expected distributions. Hence, the range of possible RMSEs is just a smaller subinterval of the range yielded by re-ratings. However, the fundamental problem can not be solved even with the pdf-rating, since very large overlaps are still possible. The obvious way of reducing the measurement uncertainty is to reduce the length of confidence intervals that scale with $1/\sqrt{N}$. Thus, the larger our sample, the smaller the intervals and the borderline cases of the RMSE will converge into a stationary state. 
By freezing the point estimators of all rating-distributions while artificially increasing the sample size, we may estimate the necessary amount of ratings to enforce convergence, so we can speak of the \textit{true} RMSE of a recommender system.
When opting for the intersection area of the minimum and maximum RMSE as a measure of this convergence, we may show the necessity of 1000-2000 ratings in order to gain intersections of more than 90\%, i.e. the minimum and maximum RMSE become mostly equivalent. This means that every user in a real rating scenario would have to re-evaluate the same item at least 1000 times to locate the RMSE-distribution accurately. However, when using the pdf-rating proceeding, a particular user would yet still have to repeat the rating task at least 40 times.

\begin{figure}[b]
\centering
\includegraphics[width=.4\textwidth]{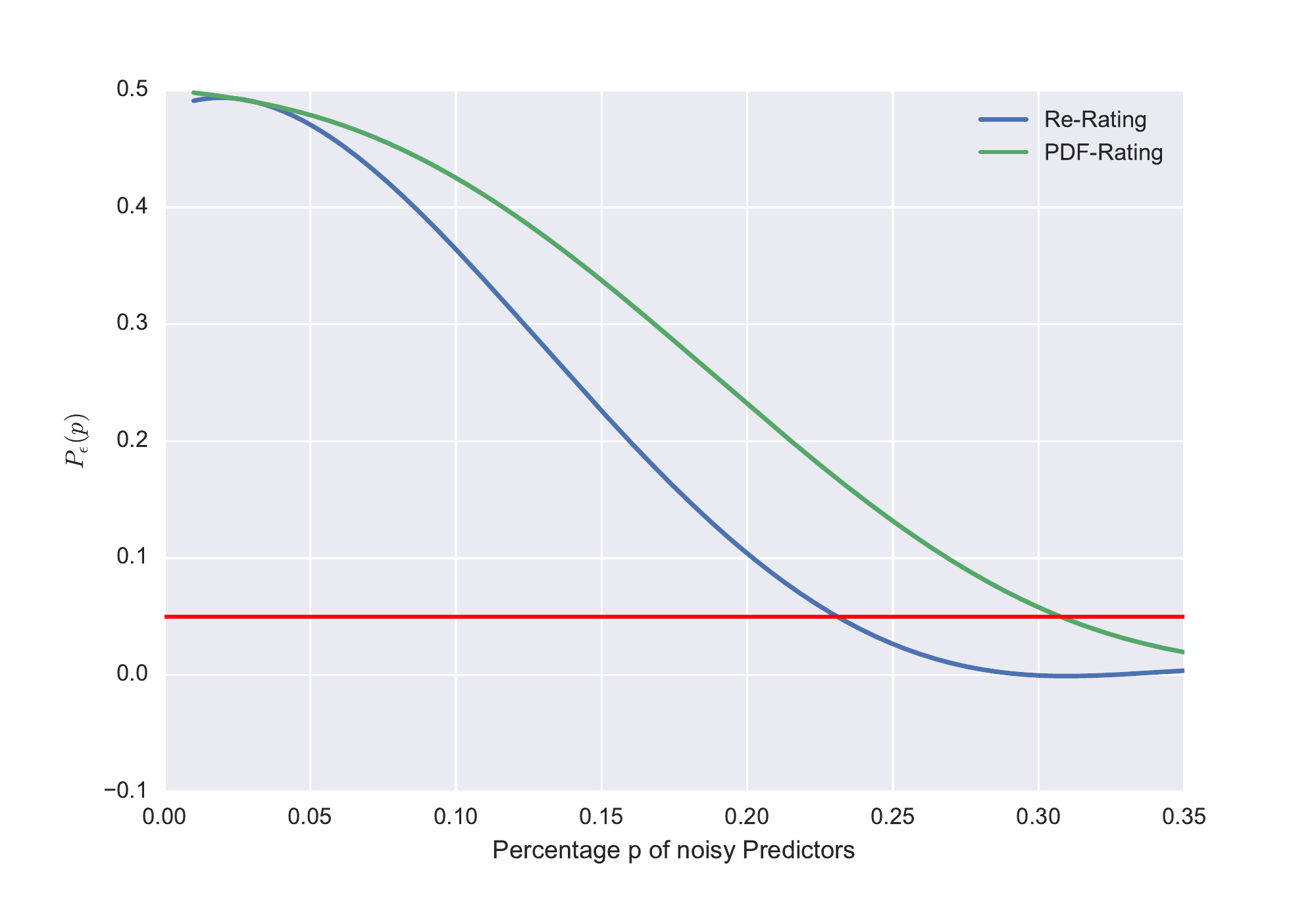}
\caption{Error Probabilities for the ranking $X_{opt}<X_{noise}$ using the re-rating and pdf-rating}
\label{fig:Noise}
\end{figure}

So far we have only considered the effects of measurement uncertainty on a single RMSE-distribution.
However, human uncertainty (associated to the width of the RMSE-distribution) leads to a much more fundamental problem, which is invariant under change of methods, namely the distinguishability of different recommender systems.
Figure \ref{fig:Noise} shows the curve of the error probability for the ranking $X_{opt}<X_{noise}$ where $X_{opt}$ is the optimal recommender and $X_{noise}$ its noise-distorted copy, whose prediction quality is worse by design.
We can see that the error probability for the re-rating drops below the 5\% mark much earlier than the error probability for the pdf-rating. Only after passing this mark, distinctions to the optimum can be considered to be significant. It is apparent that by means of the re-rating, smaller differences can be detected significantly in contrast to the pdf-rating.

\section{Discussion and Conclusions}

\paragraph*{Discussion}
Within our experiments, the normal distribution appears to be a good data model, for it is easy to handle, widely used in cognitive science for description of human properties and can not be rejected in more than 98\% of our data records.

When it comes to a comparison of the re-rating against the pdf-rating, a careful analysis of the individual distribution parameters show that the respective differences are only significant for less than 20\% of the data. Overall, the rating-distributions may have descriptive deviations, but all lie within the range of random fluctuations.
Unfortunately, each distribution is not unique, since we can only assign the parameters within confidence intervals due to a finite sample size (precision) and simulations on this basis would have to sample those parameters from their intervals (leading to reliability concerns). The comparison of confidence intervals shows that both measument approaches, model more precisely about half of the data set. The improvements of the pdf-rating for one half of this data is very large compared to the deterioration of the other half. So, on average a slight gain in overall precision occurs when using the pdf-rating. However, there are hardly any significant differences between the two methods for the rating-distributions.

The picture changes, when we consider the distributions of the variances. The actual ratings here lead to power-law-distributions, i.e. only a few people got a high extent of uncertainty while many people have very little uncertainty.
This is contrary to our everyday experience and might be an artefact of the rating instrument in which users are forced to make a discrete decision and don't have the possibility to allow other options to a certain degree. On the other hand, the distributions obtained by the pdf-rating, are normally distributed and always possess the same expectation for the ISN.
Accordingly, all users show a common uncertainty on average while each individual is scattering more or less. This essentially points to an immanent cognition, which has often been addressed specifically in our experimental setting. In case of reliability testing, the pdf-rating also leads to a higher precision, which increases sharply for higher percentiles. Overall, there is a strong deviation of the results of both measurement methods within the noise distributions. This indicates that the perceived uncertainty as an operationalisation for cognitive uncertainty is tremendously different from the uncertainty that is actually measurable through actions. This suggests that users are not able to tell us what is really going on in their minds by answering on commonly used scales.

Concerning the RMSE, both measurement approaches will lead to significant chances of the density location whereas the density spread retains. This invariance of spread under measurement approaches indicates, that the impact of human uncertainty on composed quantities can be addressed separately from the measurement method.
But nevertheless, the measurement uncertainty leads to an ambiguity of the RMSE-distribution. The range of possibilities in our example range from a total overlapping (no distinguishing at all) to very good distinguishability. The superiority of the pdf-rating in terms of precision can be observed in the range of possible outcomes. This range is just a smaller subinterval of the range yielded from re-ratings. Anyway, the fundamental problem of identifying the true distribution remains unsolved, since there are still very large intersections in the worst case. This is because the precision is set up by a frequentist translation of the entered user confidence and thus we can only reach a maximum sample size of $N = 25$. Although this is five times larger than the underlying sample size of the re-rating-proceeding, it is still far from the barrier of $N=1000$ for which the RMSE converges into a stationary state.

Our analysis of distinguishing between the RMSE of an optimal recommender and the those of a distorted copy reveals that by re-rating, lower differences can be detected significantly. In this case, the precision of the pdf-rating cannot impact this simulation because the expected distributions (stationary states) were assumed. However, both approaches have in common that the fundamental problem of distinctness still exists. This means that a recommender can only be distinguished from a supposedly better one to a certain limit, i.e. there is a natural barrier from which beyond there is only one equivalence class at good recommenders and rankings are no longer possible. This is the first statistically sound proof for the so far only as a theoretical quantity existing Magic Barrier \citep{MagicBarrier}.

It is striking that the approach which provides a location of the true state more accurately, will lead to significant distinctions only in the case of larger differences. On the other hand, the approach which detects significant distances for smaller deviations, does not allow the true RMSE to be located at all. However, both properties - (1) limiting the possibilities for the true state of two RMSEs and (2) distinguishing them - are important in real applications. Accordingly, none of the presented methods can solve both meshing problems simultaneously.

In the end, there is still the question of which specific measurement approach fits to a specific situation.
It has been shown that the distinguishability problem of a composed quantity is invariant under the measuring method and the gain in precision only marginally limits the possible states of its density. Thus, the choice of a specific approach does not matter. Likewise, the rating-distributions retrieved from both methodologies do not differ significantly.
Therefore, if analyses are carried out directly on the rating-distributions (e.g. when clustering in collaborative filtering is operated on the basis of density intersection as a measure for the similarity of two ratings), then the selection of a specific measurement method is also irrelevant. Blatant deviations do arise in the consideration of variances.
Hence, if an explicit consideration of the uncertainty is in the focus of analysis (e.g. providing additional products, search results, etc. which the user might like), then the selection of adequate measuring method is crucial. Here, the actual choice for a particular approach depends on whether human actions or human decisions shall be used for the analysis.

\paragraph*{Conclusion}
What are the consequences for user modelling and predictive data mining in general?
The essence of our contribution is the revelation of the following problems:
\begin{enumerate}
\item People are not able to tell us what they really mean.
\item Human uncertainty affects the evidence of data analysis.
\item Human uncertainty can not be measured exactly.
\end{enumerate}
At this point it must be said that these problems are not grounded in this new perspective presented here, but have always been present in data analysis. The approaches used in this contribution are just able to make these problems visible for the first time. Furthermore, these problems do not only occur in this special example, but have also been proven by us in other situations of user feedback before. These problems are therefore likely to affect any area of computer science where user feedback has to be worked on. In particular, the field of user modelling, personalisation and adaptation is strongly impacted. 
For this reason, it becomes crucial to examine the extent of impact of human uncertainty and measurement uncertainty in other situations within this field of research. It is also necessary to find proper solutions for these problems in order to keep our systems optimally adapted to human beings, i.e. not to a priori exclude appropriate possibilities and making decisions on the basis of perhaps inadequate statistical analyses. We will continue to address these issues in further research.

\nocite{*}


\vfill\eject
\bibliographystyle{ACM-Reference-Format}
\bibliography{Literature} 


\begin{thebibliography}{00}


\ifx \showCODEN    \undefined \def \showCODEN     #1{\unskip}     \fi
\ifx \showDOI      \undefined \def \showDOI       #1{{\tt DOI:}\penalty0{#1}\ }
  \fi
\ifx \showISBNx    \undefined \def \showISBNx     #1{\unskip}     \fi
\ifx \showISBNxiii \undefined \def \showISBNxiii  #1{\unskip}     \fi
\ifx \showISSN     \undefined \def \showISSN      #1{\unskip}     \fi
\ifx \showLCCN     \undefined \def \showLCCN      #1{\unskip}     \fi
\ifx \shownote     \undefined \def \shownote      #1{#1}          \fi
\ifx \showarticletitle \undefined \def \showarticletitle #1{#1}   \fi
\ifx \showURL      \undefined \def \showURL       #1{#1}          \fi
\providecommand\bibfield[2]{#2}
\providecommand\bibinfo[2]{#2}
\providecommand\natexlab[1]{#1}
\providecommand\showeprint[2][]{arXiv:#2}

\bibitem[\protect\citeauthoryear{Amatriain}{Amatriain}{2012}]%
        {workshop12}
\bibfield{editor}{\bibinfo{person}{Xavier Amatriain}} (Ed.).
  \bibinfo{year}{2012}\natexlab{}.
\newblock \bibinfo{booktitle}{{\em Workshop on Recommendation Utitlity
  Evaluation: Beyond RMSE September}}. ACM, \bibinfo{address}{Dublin, Ireland}.
\newblock


\bibitem[\protect\citeauthoryear{Amatriain and Pujol}{Amatriain and
  Pujol}{2009}]%
        {RateAgain}
\bibfield{author}{\bibinfo{person}{Xavier Amatriain} {and}
  \bibinfo{person}{Josep Pujol}.} \bibinfo{year}{2009}\natexlab{}.
\newblock \showarticletitle{Rate It Again: Increasing Recommendation Accuracy
  by User Re-rating}. In \bibinfo{booktitle}{{\em Proceedings of the Third ACM
  Conference on Recommender Systems}}. ACM, \bibinfo{pages}{173--180}.
\newblock


\bibitem[\protect\citeauthoryear{Amatriain, Pujol, and Oliver}{Amatriain
  et~al\mbox{.}}{2009}]%
        {noise1}
\bibfield{author}{\bibinfo{person}{Xavier Amatriain}, \bibinfo{person}{Josep~M.
  Pujol}, {and} \bibinfo{person}{Nuria Oliver}.}
  \bibinfo{year}{2009}\natexlab{}.
\newblock \showarticletitle{I Like It... I Like It Not: Evaluating User Ratings
  Noise in Recommender Systems}. In \bibinfo{booktitle}{{\em 17th International
  Conference on User Modeling, Adaptation, and Personalization: Formerly UM and
  AH}} {\em (\bibinfo{series}{UMAP '09})}. \bibinfo{pages}{247--258}.
\newblock


\bibitem[\protect\citeauthoryear{Beel, Genzmehr, Langer, N\"{u}rnberger, and
  Gipp}{Beel et~al\mbox{.}}{2013}]%
        {noise2}
\bibfield{author}{\bibinfo{person}{Joeran Beel}, \bibinfo{person}{Marcel
  Genzmehr}, \bibinfo{person}{Stefan Langer}, \bibinfo{person}{Andreas
  N\"{u}rnberger}, {and} \bibinfo{person}{Bela Gipp}.}
  \bibinfo{year}{2013}\natexlab{}.
\newblock \showarticletitle{A Comparative Analysis of Offline and Online
  Evaluations and Discussion of Research Paper Recommender System Evaluation}.
  In \bibinfo{booktitle}{{\em Proceedings of the International Workshop on
  Reproducibility and Replication in Recommender Systems Evaluation}} {\em
  (\bibinfo{series}{RecSys '13})}. \bibinfo{pages}{7--14}.
\newblock


\bibitem[\protect\citeauthoryear{Buffler, Allie, and Lubben}{Buffler
  et~al\mbox{.}}{2001}]%
        {Buffler}
\bibfield{author}{\bibinfo{person}{Andy Buffler}, \bibinfo{person}{Saalih
  Allie}, {and} \bibinfo{person}{Fred Lubben}.}
  \bibinfo{year}{2001}\natexlab{}.
\newblock \showarticletitle{The development of first year physics students'
  ideas about measurement in terms of point and set paradigms}.
\newblock \bibinfo{journal}{{\em International Journal of Science Education\/}}
  \bibinfo{volume}{23}, \bibinfo{number}{11} (\bibinfo{year}{2001}),
  \bibinfo{pages}{1137--1156}.
\newblock


\bibitem[\protect\citeauthoryear{Chan}{Chan}{2011}]%
        {Chan}
\bibfield{author}{\bibinfo{person}{F~Kenneth Chan}.}
  \bibinfo{year}{2011}\natexlab{}.
\newblock \showarticletitle{Miss Distance--Generalized Variance Non-Central Chi
  Distribution}. In \bibinfo{booktitle}{{\em AAS/AIAA Space Flight Mechanics
  Meeting}}. \bibinfo{pages}{11--175}.
\newblock


\bibitem[\protect\citeauthoryear{Grabe}{Grabe}{2011}]%
        {Grabe}
\bibfield{author}{\bibinfo{person}{Michael Grabe}.}
  \bibinfo{year}{2011}\natexlab{}.
\newblock \bibinfo{booktitle}{{\em Grundriss der Generalisierten Gau{\ss}'schen
  Fehlerrechnung}}.
\newblock \bibinfo{publisher}{Springer Berlin Heidelberg}.
\newblock


\bibitem[\protect\citeauthoryear{Heinicke and Jasberg}{Heinicke and
  Jasberg}{2015}]%
        {HeinickeJasberg}
\bibfield{author}{\bibinfo{person}{Susanne Heinicke} {and}
  \bibinfo{person}{Kevin Jasberg}.} \bibinfo{year}{2015}\natexlab{}.
\newblock \showarticletitle{Learning About Measurement Uncertainty in an
  Alternative Approach to Traditional Error Calculation}. In
  \bibinfo{booktitle}{{\em Electronic Proceedings of the ESERA 2015
  Conference}}, Vol.~\bibinfo{volume}{Part 1: Learning science: conceptual
  understanding}. \bibinfo{pages}{p. 265--270}.
\newblock


\bibitem[\protect\citeauthoryear{Herlocker}{Herlocker}{2004}]%
        {Herlocker}
\bibfield{author}{\bibinfo{person}{Herlocker}.}
  \bibinfo{year}{2004}\natexlab{}.
\newblock \showarticletitle{Evaluating collaborative filtering recommender
  systems}.
\newblock \bibinfo{journal}{{\em ACM Transactions on Information Systems\/}}
  \bibinfo{volume}{22}, \bibinfo{number}{1} (\bibinfo{year}{2004}),
  \bibinfo{pages}{5--53}.
\newblock


\bibitem[\protect\citeauthoryear{Iannario}{Iannario}{2014}]%
        {cub}
\bibfield{author}{\bibinfo{person}{Maria Iannario}.}
  \bibinfo{year}{2014}\natexlab{}.
\newblock \showarticletitle{Modelling Uncertainty and Overdispersion in Ordinal
  Data}.
\newblock \bibinfo{journal}{{\em Communications in Statistics - Theory and
  Methods\/}}  \bibinfo{volume}{43} (\bibinfo{year}{2014}),
  \bibinfo{pages}{771--786}.
\newblock
Issue 14.


\bibitem[\protect\citeauthoryear{JCGM}{JCGM}{2008a}]%
        {GUM}
\bibfield{author}{\bibinfo{person}{JCGM}.} \bibinfo{year}{2008}\natexlab{a}.
\newblock \bibinfo{booktitle}{{\em Guide to the Expression of Uncertainty in
  Measurement}}.
\newblock \bibinfo{type}{{T}echnical {R}eport}. \bibinfo{institution}{BIPM}.
\newblock


\bibitem[\protect\citeauthoryear{JCGM}{JCGM}{2008b}]%
        {GUMsupp1}
\bibfield{author}{\bibinfo{person}{JCGM}.} \bibinfo{year}{2008}\natexlab{b}.
\newblock \bibinfo{booktitle}{{\em Supplement 1 to the GUM - Propagation of
  distributions using a Monte Carlo method}}.
\newblock \bibinfo{type}{{T}echnical {R}eport}. \bibinfo{institution}{BIPM}.
\newblock


\bibitem[\protect\citeauthoryear{Nguyen}{Nguyen}{2013}]%
        {NewInterfaces}
\bibfield{author}{\bibinfo{person}{Tien~et.al. Nguyen}.}
  \bibinfo{year}{2013}\natexlab{}.
\newblock \showarticletitle{Rating Support Interfaces to Improve User
  Experience and Recommender Accuracy}. In \bibinfo{booktitle}{{\em Proceedings
  of the 7th ACM Conference on Recommender Systems}}. ACM,
  \bibinfo{pages}{149--156}.
\newblock


\bibitem[\protect\citeauthoryear{Peck and Devore}{Peck and Devore}{2017}]%
        {VarConfInt}
\bibfield{author}{\bibinfo{person}{Peck} {and} \bibinfo{person}{Jay~L.
  Devore}.} \bibinfo{year}{2017}\natexlab{}.
\newblock \bibinfo{booktitle}{{\em Statistics: The Exploration and Analysis of
  Data}}.
\newblock \bibinfo{publisher}{Brooks / Cole}.
\newblock


\bibitem[\protect\citeauthoryear{Ricci, Rokach, and Shapira}{Ricci
  et~al\mbox{.}}{2015}]%
        {handbook}
\bibfield{author}{\bibinfo{person}{Francesco Ricci}, \bibinfo{person}{Lior
  Rokach}, {and} \bibinfo{person}{Bracha Shapira}.}
  \bibinfo{year}{2015}\natexlab{}.
\newblock \bibinfo{booktitle}{{\em Recommender Systems Handbook}}.
\newblock \bibinfo{publisher}{Springer}.
\newblock


\bibitem[\protect\citeauthoryear{Said and Bellog{\'\i}n}{Said and
  Bellog{\'\i}n}{2014}]%
        {notConsistent}
\bibfield{author}{\bibinfo{person}{Alan Said} {and} \bibinfo{person}{Alejandro
  Bellog{\'\i}n}.} \bibinfo{year}{2014}\natexlab{}.
\newblock \showarticletitle{Comparative recommender system evaluation:
  benchmarking recommendation frameworks}. In \bibinfo{booktitle}{{\em
  Proceedings of the 8th ACM Conference on Recommender systems}}. ACM,
  \bibinfo{pages}{129--136}.
\newblock


\bibitem[\protect\citeauthoryear{Said, Jain, Narr, and Plumbaum}{Said
  et~al\mbox{.}}{2012}]%
        {MagicBarrier}
\bibfield{author}{\bibinfo{person}{Alan Said}, \bibinfo{person}{Brijnesh Jain},
  \bibinfo{person}{Sascha Narr}, {and} \bibinfo{person}{Till Plumbaum}.}
  \bibinfo{year}{2012}\natexlab{}.
\newblock \showarticletitle{Users and Noise: The Magic Barrier of Recommender
  Systems}.
\newblock In \bibinfo{booktitle}{{\em User Modeling, Adaptation, and
  Personalization}}. Vol.~\bibinfo{volume}{7379}. \bibinfo{publisher}{Springer
  Berlin / Heidelberg}, \bibinfo{pages}{237--248}.
\newblock


\bibitem[\protect\citeauthoryear{Said and Tikk}{Said and Tikk}{2012}]%
        {3DBenchmark}
\bibfield{author}{\bibinfo{person}{Alan Said} {and} \bibinfo{person}{Domonkos
  Tikk}.} \bibinfo{year}{2012}\natexlab{}.
\newblock \showarticletitle{Recommender systems evaluation: A 3D benchmark}. In
  \bibinfo{booktitle}{{\em ACM RecSys 2012 workshop on Recommendation utility
  evaluation: beyond RMSE, Dublin, Ireland}}. ACM, \bibinfo{pages}{21--23}.
\newblock


\bibitem[\protect\citeauthoryear{Sizov}{Sizov}{2016}]%
        {SizovPoster}
\bibfield{author}{\bibinfo{person}{Sergej Sizov}.}
  \bibinfo{year}{2016}\natexlab{}.
\newblock \showarticletitle{Assessment of Rating Prediction under Response
  Uncertainty}. In \bibinfo{booktitle}{{\em ACM Conference on Web Science}}.
\newblock


\end{thebibliography}

\end{document}